\documentclass[useAMS,usenatbib]{mn2e}
\newif\ifAMStwofonts

\def\pmb#1{\mbox{\boldmath$#1$}}
\def\gtsim {>\kern-1.2em\lower1.1ex\hbox{$\sim$}}
\def\ltsim {<\kern-1.2em\lower1.1ex\hbox{$\sim$}}
\def\gtsim {>\kern-1.2em\lower1.1ex\hbox{$\sim$}}
\def\ltsim {<\kern-1.2em\lower1.1ex\hbox{$\sim$}}

\def\be{\begin{equation}}
\def\ee{\end{equation}}

\usepackage{epsfig}
\begin{document}

\title{Stability of $g$-modes in rotating B-type stars}
\author[Aprilia, U. Lee, and  H. Saio]{Aprilia$^1$\thanks{E-mail: aprilia@astr.tohoku.ac.jp},
 Umin Lee$^1$\thanks{E-mail: lee@astr.tohoku.ac.jp}, and 
Hideyuki Saio$^1$\thanks{E-mail: saio@astr.tohoku.ac.jp}
\\$^1$Astronomical Institute, Tohoku University, Sendai, Miyagi 980-8578, Japan}

\date{Typeset \today ; Received / Accepted}
\maketitle


\begin{abstract}
We have studied the stability of low degree  $g$-modes in uniformly rotating B-type stars, 
taking into account 
the effects of the Coriolis force and the rotational deformation. 
From an analysis treating rotation frequency as a small parameter
it is found that slow rotation tends to {\it destabilize} high radial-order {\it retrograde}
$g$-modes, although the effect is very small or absent for relatively low order modes.
Calculating eigenfrequencies at selected rotation rates,
we find, on the other hand, that rapid rotation 
tends to {\it stabilize} {\it retrograde} $g$-modes.
The stabilizing effect appears stronger for less massive B-type stars having low effective
temperatures.
If we change rotation rate continuously, the frequency of a $g$-mode belonging to ($\ell,m$)
crosses frequencies of other $g$-modes belonging to ($\ell',m$).
If the parity of the two encountering modes are the same, they interact
each other and the stability (i.e., imaginary part of eigenfrequency) 
of each mode is  modified.
Using an asymptotic method 
we discuss the property of such mode crossings and couplings. 
For rapidly rotating stars mode couplings are important for
the stability of low degree $g$-modes. In particular, we find that the stabilization of 
retrograde $g$-modes in rapidly rotating stars is due to many strong mode 
couplings,
while {\it prograde sectoral} modes are exceptionally
immune to the damping effects from the mode couplings.
\end{abstract}


\section{Introduction}
The kappa-mechanism associated with the iron opacity bump at $T\sim2\times10^5$K 
excites high-order g-modes in intermediate-mass ($7\ga M/M_\odot \ga 3$) main-sequence
stars \citep{gs93,dmp93}. 
These B-type g-mode pulsators with typical periods of the order of days 
are called slowly pulsating B (SPB) stars, which \citet{wa91} first recognized as an 
independent class of variables \citep[see][for a review]{dec07}.
It is generally considered that majority of SPB stars are slow rotators 
with projected equatorial rotation velocities of $V\sin i \la 100$~km~s$^{-1}$, 
although several SPB stars are known to have $V\sin i$ exceeding 200~km~s$^{-1}$.
However, the criterion for slow rotation may not be appropriate   
when we look into the effect of rotation on g-modes of SPB stars.
For example, an equatorial rotation velocity of 50~km~s$^{-1}$ corresponds to 
an angular velocity $\Omega = 0.1\sqrt{GM/R^3}$
if we adopt mass and radius typical for a SPB star; 
$M=4M_\odot$ and radius, $R\approx2.4R_\odot$, 
where $G$ is the gravitational constant.
This indicates that the centrifugal acceleration, $R\Omega^2$ is much smaller than 
the surface gravity, $GM/R^2$, causing negligible deformation of the structure of the star.
However, the 50km~s$^{-1}$ velocity corresponds to a rotation period of about 2.4days,
comparable to the typical pulsation periods of SPB stars; in other words, the ratio of 
rotation to pulsation frequency is comparable to unity.
Since the Coriolis force affects stellar oscillations significantly when
$2\Omega/\omega \ga 1$ with $\omega$ being the oscillation frequency in the
co-rotating frame \citep[e.g.,][]{unno89}, g-modes in most SPB stars might be affected significantly
by the Coriolis force, even though rotational deformations are negligibly small.

In addition, it is known that $g$-modes are present in many rapidly rotating Be
stars \citep[e.g.,][]{riv03}. This is understandable because many Be stars
are located in the SPB instability strip on the HR diagram.
There is some evidence that the excitation of $g$-modes in Be
stars should be affected significantly by rotation \citep[e.g.,][]{wal05,sa07}.
Therefore, it is important to examine the effect of rotation on  $g$-modes in B-type stars.

Stability analyses of $g$-modes in rotating B stars have been carried out by several authors 
\citep[e.g.,][]{Lee01,tow05,sav05,wal05,sa07,cam08}.
\citet{Lee01} calculated
non-adiabatic $g$-modes of rotating SPB stars by taking into account  the effect of
the Coriolis force 
and found that the low degree unstable {\it retrograde} $g$-modes in the absence of rotation
are stabilized by
rapid rotation via mode couplings with high $\ell$ stable $g$-modes.
\citet{sav05} obtained similar results on the stabilization of retrograde modes 
in rapidly rotating models, by
solving a set of two dimensional partial differential equations describing 
small amplitude non-adiabatic oscillations
of uniformly rotating stars. 
The tendency of retrograde modes being stabilized is more pronounced
for g-modes in nearly critically rotating Be stars 
\citep{wal05,sa07,cam08}.  

On the other hand, such stabilization of retrograde modes does not occur
if the traditional approximation is employed \citep{tow05}. 
The horizontal component of the angular velocity of rotation ($\Omega\sin\theta$ with
$\theta$ being co-latitude) is neglected in the traditional approximation,
which makes the set of the governing equations similar to the one without rotation.
Only difference is that $\ell(\ell+1)$ in the non-rotating case is replaced with
$\lambda_{km}$ (a function of $2\Omega/\omega$) in the traditional approximation,
where $k=0, 2, 4,\ldots$ for even modes, while $k=1,3,5, \ldots$ for odd modes,
in parallel to $\ell=|m|+k$ in the non-rotating case.
The traditional approximation works well for g-modes except when
two modes associated with the same $m$ and parity (even or odd) 
but different $\lambda_{km}$s have
similar frequencies; in the traditional approximation the two modes are independent,
while the two modes couple in reality \citep{Lee89}. 
Therefore, the difference in the stability result based on the traditional approximation
from the result without it can be explained by the absence of mode couplings
under the traditional approximation. 
This indicates that mode couplings are important in the stability of $g$-modes
in rotating stars. In this paper, we discuss the effect of the couplings on
the stability of g-modes in detail.

\section{Method of calculation}

We use the method of calculation given in \citet{Lee95} 
to study the stability of 
$g$-modes in uniformly rotating stars by taking into account
the effects of the Coriolis force and the centrifugal force.
We employ a coordinate system $(a,\theta,\phi)$ 
for a rotationally deformed star, 
where the coordinate $a$ is regarded as the mean distance of equi-potential 
surface from the stellar centre. 
It is related to spherical polar coordinates $(r,\theta,\phi)$ as
\be
r=a\left[1+\epsilon\left(a,\theta\right)\right],
\ee
where $\epsilon$, a rotational deformation, can be written as
\be
\epsilon=\alpha(a)+\beta(a)P_2(\cos\theta)
\label{eq:alpbeta}
\ee
with the second Legendre polynomial $P_2(\cos\theta)=(3\cos^2\theta-1)/2$.
The term $\alpha$ represents the spherical expansion and $\beta$ the
deformation of the star due to rotation.
Assuming that $\epsilon$ is proportional to $\Omega^2$, 
we calculate the functions $\alpha(a)$ and $\beta(a)$ 
by applying the Chandrasekhar-Milne expansion
to the hydrostatic and Poisson equations for the star
\citep[see][for details]{Lee95}.

We express the angular dependence of small amplitude oscillations of the star using
finite series expansion in terms of spherical harmonic functions $Y_l^m(\theta,\phi)$.
For a given azimuthal wavenumber $m$, 
the displacement vector $\xi(a,\theta,\phi,t)$ is given by
\be
{\xi_a\over a}=\sum_{j=1}^{j_{\rm max}}S_{l_j}(a)Y_{l_j}^me^{{\rm i}\omega t},
\label{eq:xia}
\ee
\be
{\xi_\theta\over a}=\sum_{j=1}^{j_{\rm max}}\left[H_{l_j}(a){\partial\over\partial\theta}
Y_{l_j}^m+T_{l'_j}(a){1\over\sin\theta}{\partial\over\partial\phi}Y_{l'_j}
\right]e^{{\rm i}\omega t},
\ee
\be
{\xi_\phi\over a}=\sum_{j=1}^{j_{\rm max}}\left[H_{l_j}(a){1\over\sin\theta}{\partial\over\partial\phi}
Y_{l_j}^m - T_{l'_j}(a){\partial\over\partial\theta}Y_{l'_j}
\right]e^{{\rm i}\omega t},
\ee
and the Eulerian perturbation of the pressure, $p'$, for example, is given by
\be
p'(a,\theta,\phi,t)=\sum_{j=1}^{j_{\rm max}}p'_{l_j}(a)Y_{l_j}^m(\theta,\phi)e^{{\rm i}\omega t},
\label{eq:p'}
\ee
where $\omega\equiv\sigma+m\Omega$ is the oscillation frequency in the co-rotating 
frame of the star with $\sigma$ being the oscillation frequency in an inertial frame, and
$l_j=|m|+2(j-1)$ and $l'_j=l_j+1$ for even modes, and 
$l_j=|m|+2j-1$ and $l'_j=l_j-1$ for odd modes, where $j=1,~2,~\cdots,~j_{\rm max}$, and $j_{\rm max}$
is the length of the expansions.
Note that the angular pattern of $p'$ for even (odd) modes is symmetric (antisymmetric)
with respect to the equator.
Substituting these expansions into the linearized basic equations, we obtain a finite set of
coupled first order linear ordinary differential equations for the expansion coefficients 
$S_{l_j}(a)$ and $p'_{l_j}(a)$, and for given values of the parameters $m$ and $\Omega$, we
solve the set of the differential equations as an eigenvalue problem of $\omega$ by imposing appropriate boundary conditions
at the centre and surface of the star.
The set of the linear differential equations as well as the boundary conditions
is given in \citet{Lee95}. 
Since we assume that the perturbed quantities are proportional 
to $e^{{\rm i}(m\phi+\omega t)}$,  
modes having negative $\omega_{\rm I}$ are
pulsationally unstable (or excited), 
where $\omega_{\rm I}\equiv {\rm Im}(\omega)$ and $\omega_{\rm R}\equiv {\rm Re}(\omega)$ denote respectively 
the imaginary part and the real part of $\omega$.
Note that when $\omega_{\rm R}>0$, modes associated with negative (positive) azimuthal wavenumber $m$ are prograde (retrograde) modes 
seen in the co-rotating frame of the star.

To save computing time, we employ the Cowling approximation, neglecting the Eulerian 
perturbation of the gravitational potential.  
We ignore the effect of rotational spherical expansion of the equilibrium structure 
on the oscillation, that is, we set $\alpha=0$ in equation~(\ref{eq:alpbeta}),
because it is just a change in mean radius not important for our qualitative study. 

To determine an adequate length of the series expansions, $j_{\rm max}$, 
we have compared the results 
of sample calculations obtained with $j_{\rm max}=6$, 8, 10, and 12.
We have found that $j_{\rm max}=10$ is large enough for  
unstable $g$-modes even for the case of rapid rotation $\Omega=0.4\sqrt{GM/R^3}$.
Therefore, we have adopted $j_{\rm max}=10$ in calculating eigenfrequencies 
and eigenfunctions of $g$-modes presented in this paper.

\subsection{Mode identification and notation}

\begin{figure}
\resizebox{0.45\textwidth}{!}{
\includegraphics{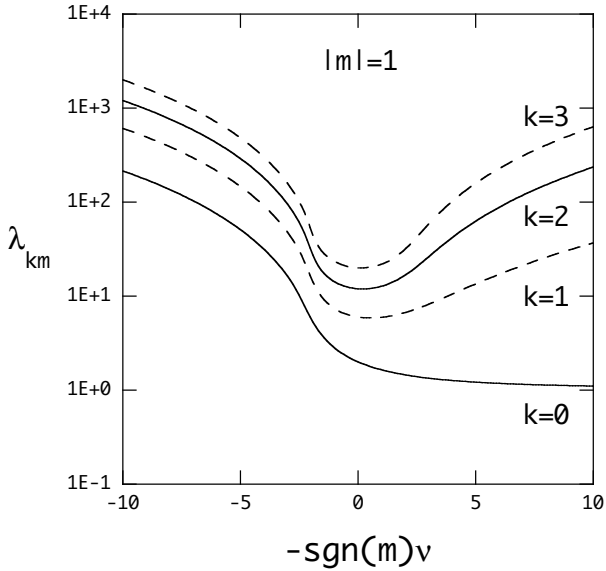}}
\caption{Eigenvalue of Laplace's tidal equation $\lambda_{km}$ versus $-{\rm sgn}(m)\nu$ 
for $|m|=1$, where ${\rm sgn}(m)\equiv m/|m|$ and $\nu\equiv 2\Omega/\omega$.
Solid lines are for even modes and dashed lines for odd modes.}
\label{fig:lambda}
\end{figure}

As equations~(\ref{eq:xia})--(\ref{eq:p'}) indicate, the angular variation of
a nonradial pulsation mode in a rotating star cannot be represented by a single
spherical harmonic but need to be expanded into many harmonics for a given
azimuthal order $m$.
The angular variation of a mode depends on the ratio of rotation frequency
to the oscillation frequency.
To distinguish the angular dependence property of each mode, 
it is useful to refer to the mode properties under the traditional approximation, 
in which the term $-\Omega\sin\theta\pmb{e}_\theta
$ in $\pmb{\Omega}=\Omega\cos\theta\pmb{e}_r
-\Omega\sin\theta\pmb{e}_\theta$ is neglected.
Many properties of low frequency oscillations of uniformly rotating stars
are well explained by using the traditional approximation 
except for mode couplings \citep[e.g.,][]{Lee87,Lee97}. 
In this approximation, the angular dependence of the oscillations are
given by Hough function \citep[e.g.,][]{Lin68} 
which is the eigenfunction, associated with
the eigenvalue $\lambda_{km}$, of Laplace's tidal equation.
For a given azimuthal wavenumber $m$, $\lambda_{km}$ 
depends on the ratio $\nu\equiv 2\Omega/\omega$
and tends to $\ell(\ell+1)$ with $\ell=|m|+k$ as $\nu\rightarrow 0$ for $k\ge0$.
Fig.~\ref{fig:lambda} shows $\lambda_{km}$ as a function of $\nu$ 
for $k=0-3$ with $|m|=1$.
The quantity  $\sqrt{\lambda_{km}}$ represents a kind of surface wave number.
Except for the prograde sectoral modes ($k=0$), $\lambda_{km}$ increases 
as the parameter $\nu$ increases.
The prograde sectoral modes (associated with $\lambda_{0m}$ $(m<0)$) 
are special modes in rapidly rotating stars whose surface wavenumber is 
lower than the value at $\Omega=0$, and hardly changes with $\Omega$. 
We emphasize here that although we use the modal properties under the 
traditional approximation as guidelines to understand the behavior of 
g-modes as function of $\Omega$, we do {\it not} use the traditional 
approximation in our numerical analyses.  

To identify calculated modes, we make use of the fractional kinetic 
energy of oscillation mode defined as
\be
f_j={e_j/ \sum_{i=1}^{j_{\rm max}}e_i},
\ee
where
\be
\begin{array}{l} \displaystyle
e_j=\int_0^R\left[\left|S_{l_j}\right|^2
+ l_j\left(l_j+1\right)\left|H_{l_j}\right|^2  \right.\cr
\displaystyle \left. \hspace{0.4\columnwidth}
+l'_j\left(l'_j+1\right)\left|T_{l'_j}\right|^2\right]4\pi\rho a^4da,
\end{array}
\ee
and $l_j=|m|+2(j-1)$ and $l'_j=l_j+1$ for even modes, and 
$l_j=|m|+2j-1$ and $l'_j=l_j-1$ for odd modes.

If we find $f_{j}\ge 0.5$ with a particular value of $j$
for a calculated {\it even} mode with a given $m$, we identify it as a mode 
associated with $\lambda_{km}$ with $k=2(j-1)$ (if the mode is an {\it odd} mode
$k=2j-1$). 
For the case of rapid rotation, however,
we sometimes need to use a somewhat relaxed criterion
given by $f_j\ge \alpha$ with $\alpha<0.5$, particularly for modes with $j>1$.
Also, we denote the mode as $(\ell,m)$ with $\ell\equiv |m| + k$ extending
the familiar notation for nonradial pulsations of a non-rotating star.
%
Note that prograde sectoral modes correspond to $\ell=-m$.

\subsection{Unperturbed models}

For unperturbed models of the stability analyses,
we have calculated main-sequence evolution models without rotation, 
using a standard stellar evolution code using the OPAL opacity \citep{ig92,ig96}.
We have adopted masses of $4 M_\odot$ and $5 M_\odot$ with initial chemical compositions 
of $(X,Z)=(0.70,0.02)$ as typical models for SPB stars.
Table \ref{tab:param} 
shows parameters of selected models.

\begin{table}
\centering
\caption{Physical parameters of the main sequence models with $X=0.7$ and $Z=0.02$}
\begin{tabular}{@{}ccccccc@{}}
\hline
&\multicolumn{3}{c}{$M=4M_\odot$} &\multicolumn{3}{c}{$M=5M_\odot$} \\
\noalign{\vskip 3pt}
Model & $\log {L\over L_\odot}$ & $\log T_{\rm eff}$ & $X_c$& $\log {L\over L_\odot}$ & $\log T_{\rm eff}$ & $X_c$\\
\hline
\noalign{\vskip 3pt}
A & 2.371 & 4.165 & 0.700 & 2.727 & 4.226 & 0.700\\
B & 2.464 & 4.144 & 0.476 & 2.825 & 4.208 & 0.481\\
C & 2.519 & 4.120 & 0.308 & 2.886 & 4.185 & 0.311\\
D & 2.565 & 4.083 & 0.105 & 2.939 & 4.149 & 0.105\\
\hline
\end{tabular}
\label{tab:param}
\end{table}

\section{$g$-modes excited at selected rotation frequencies}

\begin{figure*}
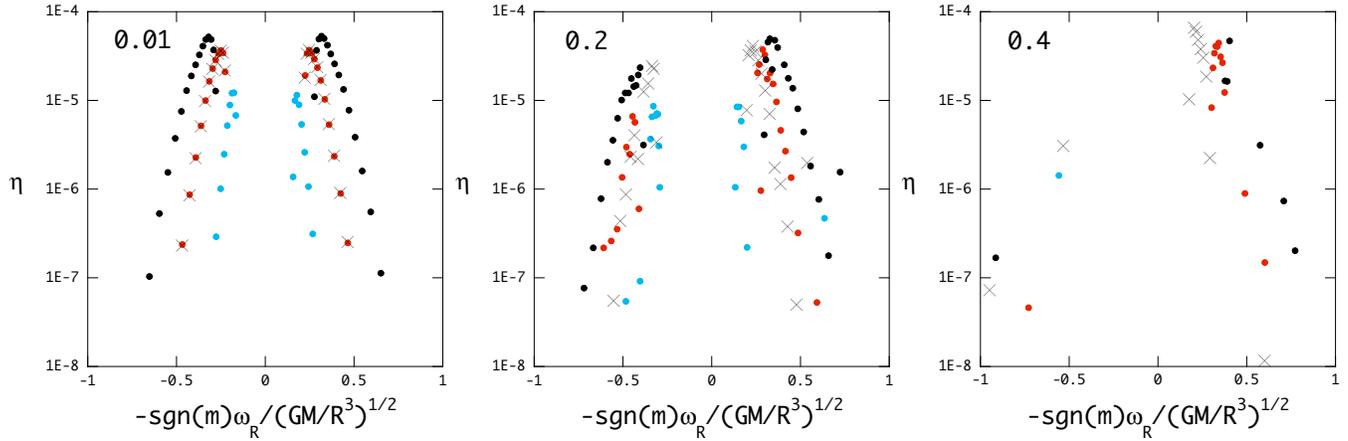

\resizebox{0.33\textwidth}{!}{
\includegraphics{f2a.epsi}}
\resizebox{0.33\textwidth}{!}{
\includegraphics{f2b.epsi}}
\resizebox{0.33\textwidth}{!}{
\includegraphics{f2c.epsi}}
\caption{Growth rate $\eta\equiv-\omega_{\rm I}/\omega_{\rm R}$ versus 
$-{\rm sgn}(m)\bar\omega_{\rm R}$
for $g$-modes excited in the $4M_\odot$ ZAMS model A, where the cyan, red, and black dots
stand for the $g$-modes with  
 $(\ell,|m|)=(1,1)$, (2,1), and (3,2),
respectively, and the crosses stand for the $g$-modes with  $(\ell,|m|)=(2,2)$.
Left-, middle- and right-panels are for rotation frequencies of 
$\bar\Omega=0.01$, 0.2 and 0.4, respectively.
Prograde modes having $m<0$ appear on the right half of each panel, while
retrograde modes ($m>0$) appear on the left side. 
}
\label{fig:grow4A}
\end{figure*}

\begin{figure*}
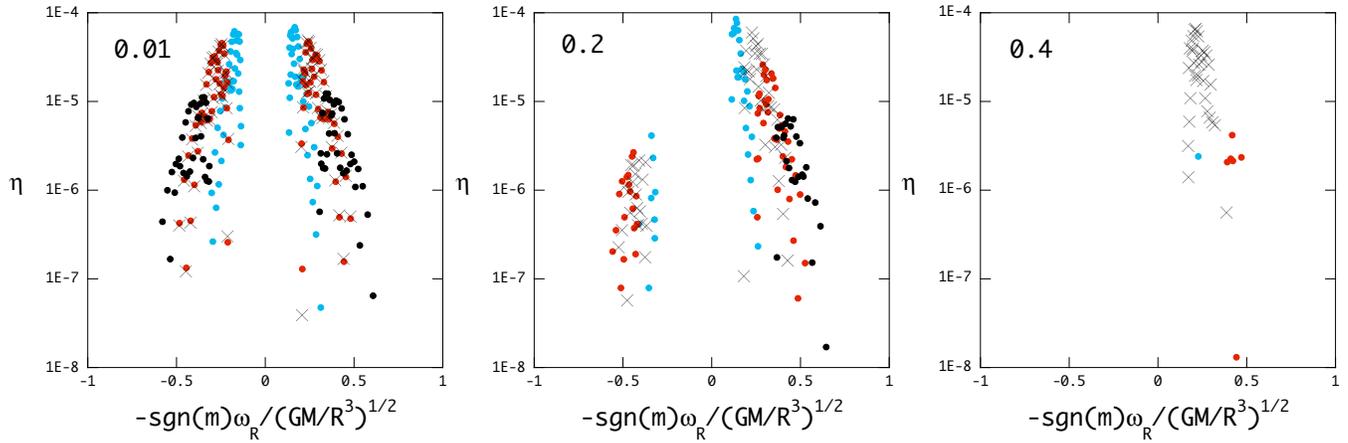

\resizebox{0.33\textwidth}{!}{
\includegraphics{f3a.epsi}}
\resizebox{0.33\textwidth}{!}{
\includegraphics{f3b.epsi}}
\resizebox{0.33\textwidth}{!}{
\includegraphics{f3c.epsi}}
\caption{Same as Fig.~\ref{fig:grow4A} but for the $4M_\odot$ evolved main sequence model D.}
\label{fig:grow4D}
\end{figure*}

\begin{figure*}
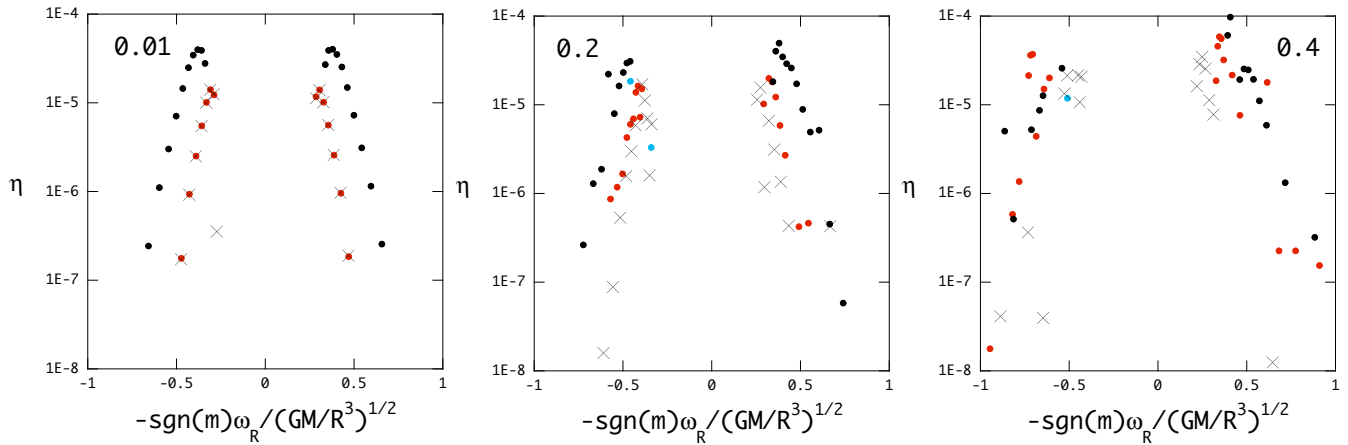

\resizebox{0.33\textwidth}{!}{
\includegraphics{f4a.epsi}}
\resizebox{0.33\textwidth}{!}{
\includegraphics{f4b.epsi}}
\resizebox{0.33\textwidth}{!}{
\includegraphics{f4c.epsi}}
\caption{Same as Fig.~\ref{fig:grow4A} but for the $5M_\odot$ ZAMS model A.}
\label{fig:grow5A}
\end{figure*}

\begin{figure*}
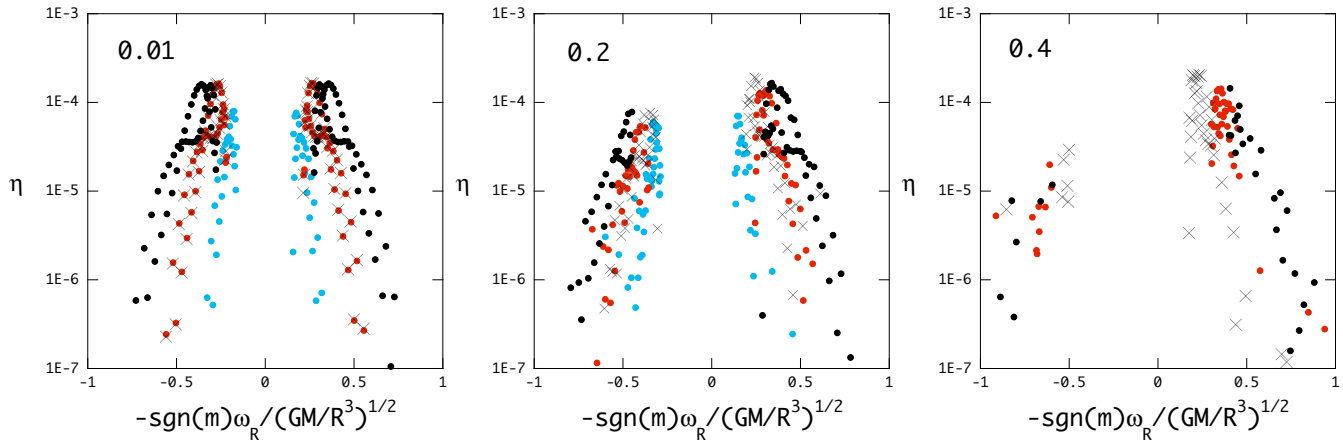

\resizebox{0.33\textwidth}{!}{
\includegraphics{f5a.epsi}}
\resizebox{0.33\textwidth}{!}{
\includegraphics{f5b.epsi}}
\resizebox{0.33\textwidth}{!}{
\includegraphics{f5c.epsi}}
\caption{Same as Fig.~\ref{fig:grow4A} but for the evolved $5M_\odot$ main sequence model D.}
\label{fig:grow5D}
\end{figure*}

Figs.~\ref{fig:grow4A}--\ref{fig:grow5D} show 
growth rates, $\eta\equiv -\omega_{\rm I}/\omega_{\rm R}$, of unstable $g$-modes
in the $4M_\odot$ and $5M_\odot$ ZAMS and D models (Tab.~\ref{tab:param})
at three selected rotation frequencies;  $\bar\Omega=0.01$, 0.2, and 0.4,
where $\bar\Omega \equiv \Omega/\sqrt{GM/R^3}$.
The horizontal axes $-{\rm sgn}(m)\bar\omega_{\rm R}$ measure the oscillation frequency 
in the co-rotating frame of the prograde (retrograde) modes in positive (negative) direction,
where $\bar\omega=\omega/\sqrt{GM/R^3}$.
Cyan and red dots are for $(\ell,|m|)=(1,1)$ and $(2,1)$, respectively, while
crosses and black dots are for $(\ell,|m|)=(2,2)$ and $(3,2)$, respectively.

Firstly let us look at the slowest rotation cases (left panels). 
When rotation is very slow, crosses and red dots are almost
superposed on each other, because eigenfrequencies are independent 
of the azimuthal order $m$  in a non-rotating star,
and the first order rotational correction to the frequency, 
which is proportional to $m$, is small for slow rotation.  
The growth rates $\eta$ as a functions of $\bar\omega_{\rm R}$ 
are almost symmetric 
with respect to the axis of $\bar\omega_{\rm R}=0$. 
Many $g$-modes are excited in a certain frequency range depending on the
effective temperature of each model; the unstable frequency range is lower
for a lower effective temperature, which is obvious, for example,  
when a ZAMS case is compared to the case of the evolved D model.
(We note that the density of g-modes is higher in the latter because 
the Brunt-V\"ais\"al\"a
frequency increases as evolution proceeds.)  
This tendency of stability arises from the optimal condition for 
the kappa-mechanism of excitation;
the kappa-mechanism is most effective when the thermal-timescale
at the Fe opacity bump is comparable to pulsation periods 
\citep[see e.g.,][]{gs93}. 
That is, longer-period modes tend to be excited in cooler models in which
the opacity bump is located in deeper layers having longer thermal-times.

In the $5 M_\odot$ ZAMS model (Fig.~\ref{fig:grow5A}), $\ell =1$ g-modes
are not excited {\bf at $\bar\Omega=0.01$}, because their frequencies 
are too low to satisfy the optimal frequency condition
for the kappa-mechanism. 
Retrograde $\ell =1$ g-modes become unstable as the rotation frequency
increases slightly as seen in the middle panel of Fig.~\ref{fig:grow5A}.
This is because the frequencies of retrograde modes have increased to
enter into the frequency range where the kappa-mechanism works strongly enough to
excite modes. 
Since the rotational effect on the frequency in a very slowly rotating star 
is approximately given as $mC_1\bar\Omega$ in the co-rotating frame with a positive
constant $C_1$ (see the next section), frequencies of retrograde modes increase as the rotation rate
increases.

It is seen in the panels for $\bar\Omega=0.2$ and 0.4 
in Figs.~\ref{fig:grow4A}--\ref{fig:grow5D}
that the prograde-retrograde
symmetry is broken as the rotation frequency increases; the maximum growth rates
for the retrograde modes become lower than those of the prograde modes, 
and the oscillation frequencies of the unstable retrograde modes increase as the rotation
frequency increases, while the effect is not appreciable for the prograde modes. 
We also note that
the number of unstable $g$-modes is largely reduced, particularly for
the retrograde $g$-modes.
This is due to selective damping caused by mode couplings, which will be
discussed in \S\ref{sec:couple}.
%

Fig.~\ref{fig:grow5D} shows unstable $g$-modes of
a $5M_\odot$ evolved main-sequence model, which has an effective temperature
similar to that of the $4M_\odot$ ZAMS model (see Table 1).
Comparing Fig.~\ref{fig:grow5D} with Fig.~\ref{fig:grow4A} for the
the $4M_\odot$ ZAMS model, we find that
although the frequency spectra for the former model are
much denser than those of the latter, 
rapid rotations stabilize retrograde modes similarly for both models.
This may suggest that the stabilizing effect on  
retrograde $g$-modes mainly depends on the 
effective temperature, while 
the difference in the internal structure due to mass difference
has only a minor effect. 

Figs.~\ref{fig:grow4A}--\ref{fig:grow5D}
clearly show that in rapidly rotating stars 
{\it retrograde} modes
tend to be stabilized more strongly than prograde modes.
In particular, for the $4 M_\odot$ evolved model shown in Fig.~\ref{fig:grow4D}
only {\it prograde} (mostly sectoral $\ell=2$) modes are excited.
The retrograde--prograde mode asymmetry is less pronounced in relatively
hotter models.
This is (at least) partly because unstable frequency ranges in hotter models
are higher (due to the optimal condition for the kappa-mechanism at
the Fe opacity bump) so that the ratios $2\Omega/\omega$ (which govern the
strength of the Coriolis force effect) for the excited modes are generally
smaller, and hence the rotation effects are weaker in hotter models.

\section{Slow rotation}

Let us next discuss the effect of slow rotation on the stability of 
$g$-modes in rotating B stars \citep[e.g.,][]{Lee01}. 
For oscillations in a slowly rotating star, we can treat the rotation 
frequency $\Omega$ as a small parameter, and
express the oscillation frequency $\omega$ of a mode as
\be
\bar\omega\left(\bar\Omega\right)=\bar\omega_0+mC_1\bar\Omega+C_2\bar\Omega^2,
\label{eq:omegaslow}
\ee
where $\omega_0$ is the oscillation frequency of the mode in the absence 
of rotation.
The coefficient $C_1$ represents the first order rotational effect due to the Coriolis force and the coefficient $C_2$ the second order effects that come from
the centrifugal force and the Coriolis force.
Since eigenfrequencies $\omega$ are complex numbers in nonadiabatic analyses,
$C_1$ and $C_2$ are also complex numbers.
The real part of $C_1$, $C_{1{\rm R}}$, is approximately equal to the adiabatic expression
\be
C_{1{\rm R}}\approx {\int_0^R[2\xi_{\rm ar}/\xi_{\rm ah}+1]\xi_{\rm ah}^2\rho r^2dr\over
\int_0^R[(\xi_{\rm ar}/\xi_{\rm ah})^2+\ell(\ell+1)]\xi_{\rm ah}^2\rho r^2dr}
\label{eq:c1ad} 
\ee
\citep[e.g.,][]{unno89}, where subscript `a' indicates adiabatic eigenfunctions.

We note that negative (positive) $C_{1{\rm I}}={\rm Im}(C_1)$
means that the Coriolis force due to a slow rotation 
has destabilizing (stabilizing) effect on 
{\it retrograde} modes with $m>0$ and stabilizing (destabilizing) 
effect on {\it prograde} modes with $m<0$,
and that negative (positive) $C_{2{\rm I}}$ means the destabilizing (stabilizing) effect on
both prograde and retrograde modes.
In this paper, we have obtained the complex coefficients $C_1$ and $C_2$ for a given $m$, 
by computing the eigenfrequencies $\bar\omega$ of a mode at three different 
rotation frequencies, e.g., $\bar\Omega=0$, $10^{-3}$ and $-10^{-3}$,
and substituting these values in equation~(\ref{eq:omegaslow}).
We confirmed that $C_1$ thus computed is in good agreement 
with that obtained with the method by \citet{car82}. 

\begin{figure*}
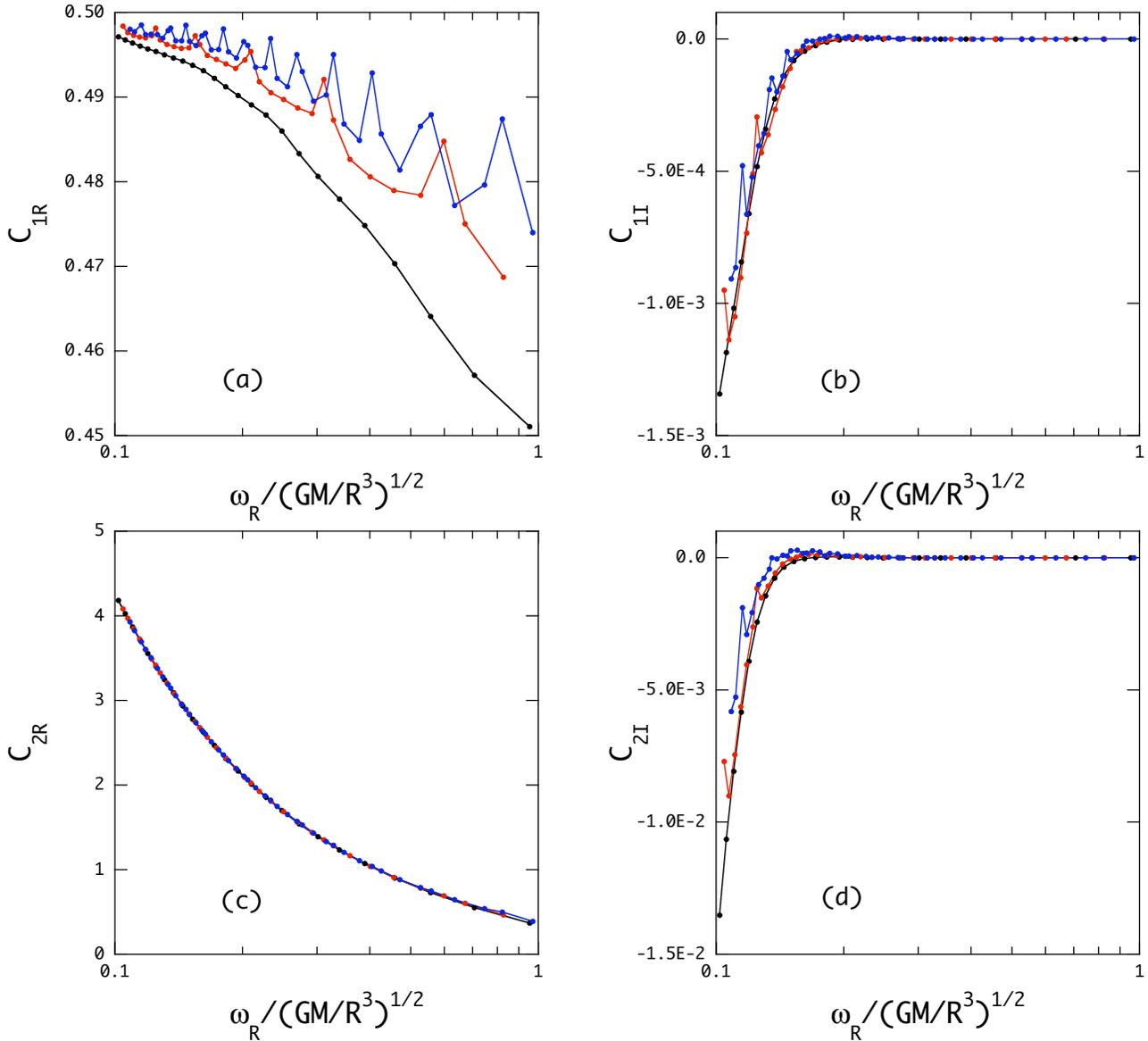

\begin{center}
\epsfig{file=f6a.epsi,width=0.45\textwidth}\hspace{0.05\textwidth}
\epsfig{file=f6b.epsi,width=0.45\textwidth}
\epsfig{file=f6c.epsi,width=0.45\textwidth}\hspace{0.05\textwidth}
\epsfig{file=f6d.epsi,width=0.45\textwidth}
\end{center}
\caption{Complex coefficients $C_{1}$ and $C_{2}$ versus $\bar\omega_{\rm R}$ for $\ell=|m|=1$ $g$-modes of
the $4M_\odot$ main sequence models, where the black, red, and blue lines indicate,
respectively, the models A (ZAMS), B, and C in Table~\ref{tab:param}.}
\label{fig:c1c2}
\end{figure*}

In Fig.~\ref{fig:c1c2}, we plot $C_1$ and $C_2$ versus $\bar\omega_{\rm R}$ 
for $\ell=|m|=1$ $g$-modes of $4M_\odot$ main sequence models, 
of which physical parameters are tabulated in Table~\ref{tab:param}.
For the ZAMS model $C_1$ and $C_2$ are essentially the same as the result of \cite{Lee01}.
The coefficients $C_{1{\rm R}}$ for the evolved models are systematically larger than that for the ZAMS model.
This can be understood as follows. In the interior of an evolved model 
the Brunt-V\"ais\"al\"a frequency $N$ is higher than in the ZAMS model.
An asymptotic theory of nonradial pulsations \citep[e.g.,][]{sh79} %
gives the ratio of radial to horizontal displacement as 
$\left|{\xi_r/ \xi_h}\right|\sim \sqrt{l(l+1)}\left|{\omega/ N}\right|$,
which indicates the ratio $|\xi_r/\xi_h|$ is smaller in the evolved models.
The smaller ratio, in turn, yields larger $C_{1R}$ as seen 
from equation~(\ref{eq:c1ad}).

In addition, $C_{1\rm R}$ for an evolved model has many 
regularly spaced
peaks. 
The peaky features are also found in the imaginary parts of $C_1$ and $C_2$ for 
the evolved models.
We have also calculated $C_1$ and $C_2$ for $10M_\odot$ main sequence models (not shown) to
find quite similar behavior of the coefficients as functions of $\bar\omega_R$.
The peaks in $C_1$ are caused by $g$-mode eigenfunctions being trapped 
in the $\mu$-gradient zone ($\mu$ stands for mean molecular weight) 
above the convective-core boundary, in which
the Brunt-V\"ais\"al\"a frequency is much higher than the adjacent layers. 
As discussed above, this yields larger $C_{1R}$ for the trapped modes.

The regularity of the peaks is also understood from the asymptotic theory, 
which gives the frequency of a $g$-mode trapped in the $\mu$-gradient zone,
$\omega_\mu$, as
\be
\omega_\mu\approx{\sqrt{l(l+1)}\over n_\mu\pi}\int_{\Delta r}{N\over r}d r, 
\label{eq:omegamu}
\ee
where $\Delta r$ and $n_\mu$ denote respectively the width 
of the $\mu$-gradient zone and the number of radial nodes within the zone.
Equation (\ref{eq:omegamu}) indicates that the frequency separation between 
two consecutive trapped $g$-modes is proportional to $[n_\mu(n_\mu+1)]^{-1}$,
which explains why peaks are reqularly spaced 
and the separations of peaks decrease with increasing
radial orders (or decreasing frequencies). 
%
%

The imaginary parts of $C_1$ and $C_2$ decrease rapidly as $\bar\omega_{\rm R}$
decreases in the range $\bar\omega_{\rm R} \la 0.3$.
Since the imaginary part of $C_1$ is negative for high radial-order $g$-modes,
slow rotation has {\it destabilizing} (stabilizing) effect 
on high radial-order {\it retrograde} (prograde) $g$-modes.
Considering that the imaginary part of $C_2$ is also negative 
for the high radial-order $g$-modes,
the destabilizing effect of slow rotation on the $g$-modes works more 
strongly for the high radial-order retrograde modes.
For relatively low radial-order modes both $C_{1{\rm I}}$ and $C_{2{\rm I}}$
are very small, indicating a slow rotation affects little the stability of
these modes. 
The steep decreases of the imaginary parts of $C_1$ and $C_2$ with decreasing 
g-mode frequencies can be understood from the relation between the period
of a high order g-mode 
and the optimal period for the kappa-mechanism excitation that is roughly
the thermal timescale at the Fe opacity bump.
The period of a very high-order g-mode that has negative 
$C_{1{\rm I}}$ and $C_{2{\rm I}}$
is much longer than the optimal period for the kappa-mechanism.
If the mode is a retrograde mode, both the first order and 
the second order effects
of rotation increase the frequency (or decrease the period) 
as eq.(\ref{eq:omegaslow}) 
indicates. This effect makes the kappa-mechanism to work stronger to the mode,
and appears as negative $C_{1{\rm I}}$ and $C_{2{\rm I}}$; in other words,
the first and the second order rotation effects tend to destabilize 
very high-order retrograde g-modes.
On the other hand, for a prograde mode the first order effect increases 
the period, while the second order effect decreases the period. 
Therefore, the first order rotation effect tends to stabilize a prograde mode,
while the second order effect tends to destabilize it.

\section{Mode couplings} 
\label{sec:couple}

\begin{figure*}
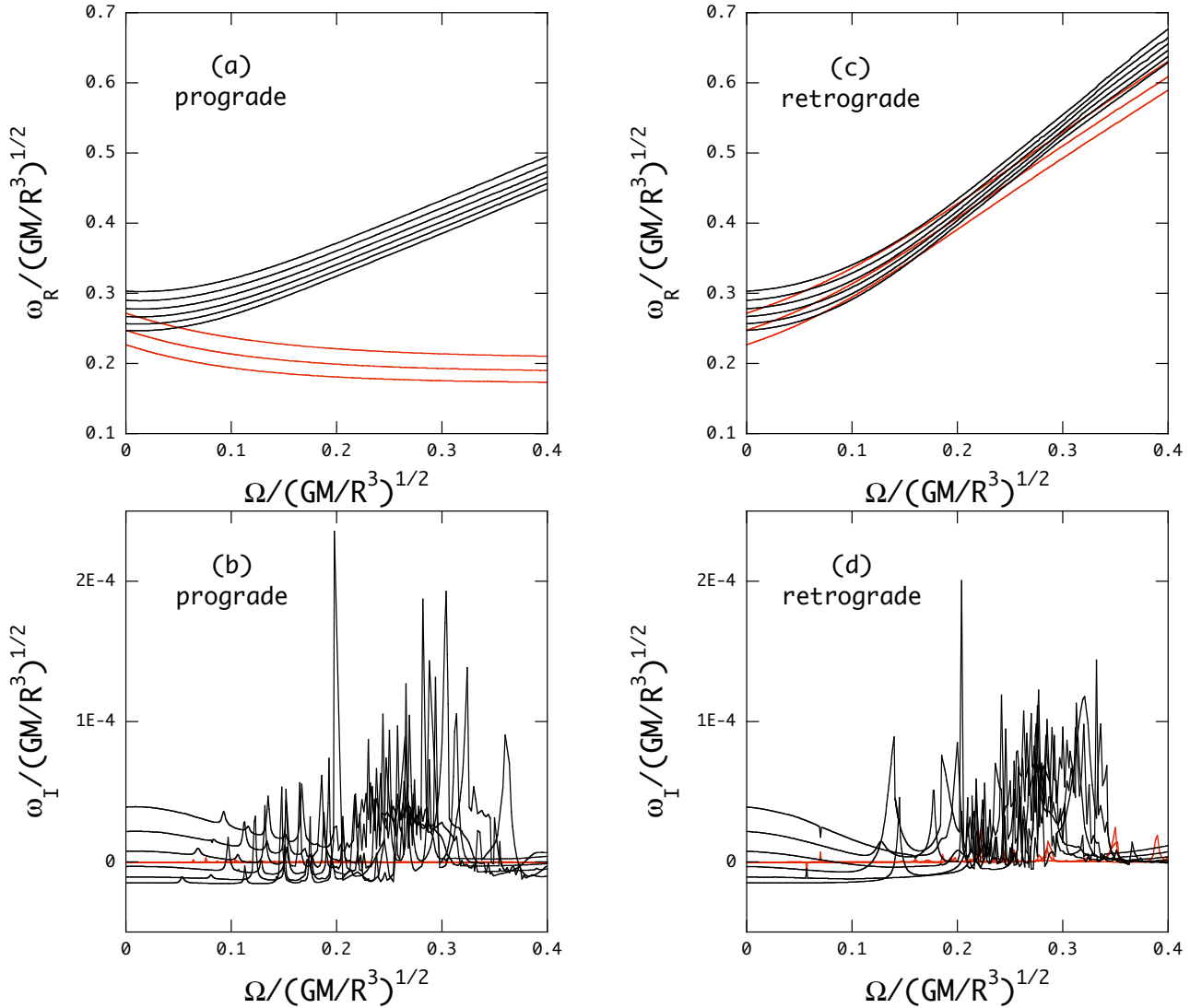

\begin{center}
\epsfig{file=f7a.epsi,width=0.45\textwidth}\hspace{0.05\textwidth}
\epsfig{file=f7c.epsi,width=0.45\textwidth}
\epsfig{file=f7b.epsi,width=0.45\textwidth}\hspace{0.05\textwidth}
\epsfig{file=f7d.epsi,width=0.45\textwidth}
\end{center}
\caption{Complex $\bar\omega$ of even $|m|=1$ $g$-modes of the $4M_\odot$ ZAMS model as functions of $\bar\Omega$ for prograde modes in
panels (a) and (b) and for retrograde modes in panels (c) and (d), where 
the red curves denote the even $\ell=1$ $g$-modes 
and the black curves the even $\ell=3$ $g$-modes.
Here, the radial order of the $g$-modes ranges from 9 to 11 
for $\ell=1$ and from 21 to 26 for $\ell=3$.}
\label{fig:omegaOmega}
\end{figure*}

\begin{figure*}
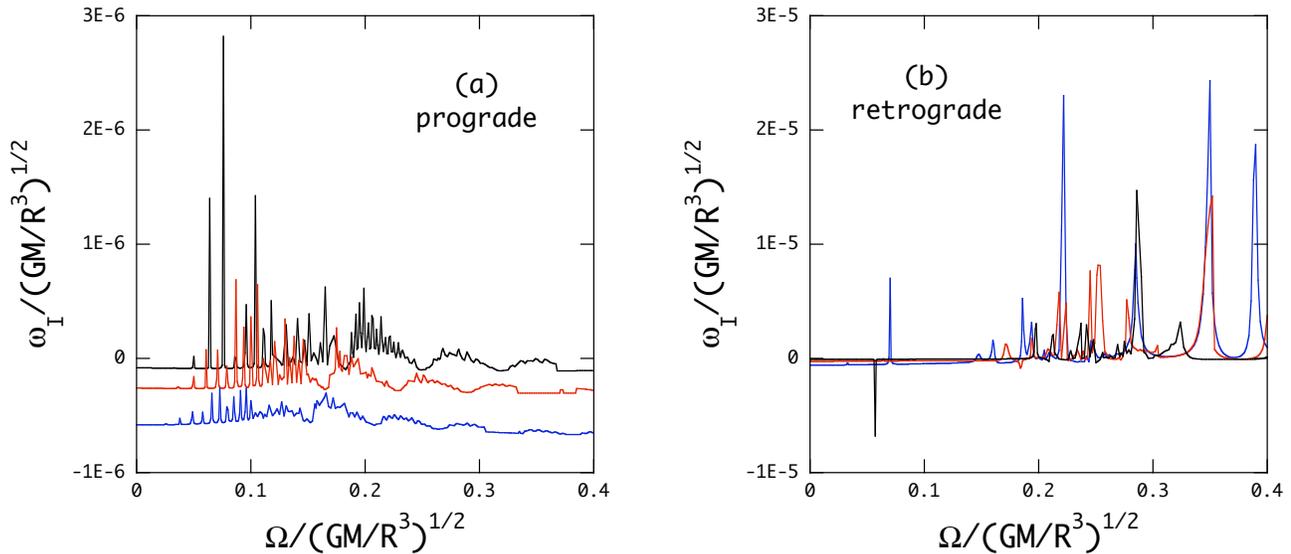

\begin{center}
\epsfig{file=f8a.epsi,width=0.45\textwidth}\hspace{0.05\textwidth}
\epsfig{file=f8b.epsi,width=0.45\textwidth}
\end{center}
\caption{Zoomed diagrams for $\bar\omega_{\rm I}$ of even $\ell=|m|=1$ $g$-modes 
of the $4M_\odot$ ZAMS model as functions of $\bar\Omega$ for prograde modes in
panel (a) and for retrograde modes in panel (b), where 
the black, red, and blue curves denote the $g_9$, $g_{10}$, and $g_{11}$-modes, respectively.}
\label{fig:omegaOmegak0}
\end{figure*}

\subsection{Numerical Results}

Fig.~\ref{fig:omegaOmega} shows the complex frequencies $\bar\omega$ of selected 
even $g$-modes of  $(\ell,|m|)=(1,1)$ (associated with $\lambda_{0\pm1}$)
(red lines) and of $(\ell,|m|)=(3,1)$ (associated with $\lambda_{2\pm1}$) (black lines) 
as functions of $\bar\Omega$ for the $4M_\odot$ ZAMS model, where 
the radial orders of the $g$-modes range from 21 to 26 
for the $\ell=3$ modes and from 9 to 11 for the $\ell=1$ modes.
The left panels, (a) and (b), are for prograde ($m=-1$) modes, while
the right panels, (c) and (d), are for retrograde ($m=+1$) modes.

The variation of pulsation frequency $\omega_{\rm R}$ of a g-mode as a function of 
$\Omega$ reflects the variation of $\lambda_{km}$ as a function of $\nu$.
For a high order g-mode, pulsation frequency is approximately written as
\be
\bar\omega_{\rm R} \approx {\sqrt{\lambda_{km}}\over n\pi}\int_0^1\bar{N} {dx\over x}, 
\label{eq:freq}
\ee
where $n$ is the radial order, $\bar{N}\equiv N/\sqrt{GM/R^3}$ normalized Brunt-V\"ais\"a\l\"a frequency,
and $x$ fractional radius $r/R$ \citep{Lee87,Lee89}.
The equation indicates that the frequency of a prograde sectoral mode ($\ell=-m=1$) 
slightly decreases but stays nearly constant as the rotation frequency increases
just as $\lambda_{0-1}$ behaves as a function of $\nu$.
The frequencies of the other modes in Fig.~\ref{fig:omegaOmega} increase at
different rates depending on the corresponding $\lambda_{km}$s. 
Therefore, the oscillation frequency 
of a $g$-mode associated with a given $\lambda_{km}$ crosses with another g-mode
associated with $\lambda_{k'm}$ ($k'\not=k$) as seen in Fig.~\ref{fig:omegaOmega}.

For adiabatic modes having real frequencies, a mode crossing between $g$-modes
leads to an avoided crossing, but for non-adiabatic modes having complex frequencies, 
mode crossing is not always an avoided one.
More importantly, a mode crossing between non-adiabatic $g$-modes can change the stability 
of the modes. 
The imaginary parts of eigenfrequencies as functions of $\Omega$ in Fig.~\ref{fig:omegaOmega}
show numerous narrow peaks large and small.
We consider that these peaks are caused by mode couplings.
In panel (d) of Fig.~\ref{fig:omegaOmega} we recognize two pairs of a small peak and 
dip around $\bar\Omega\sim 0.06$ and $\sim 0.07$, while
panel (c) shows that a black line crosses a red line at each of the same $\bar\Omega$,
indicating these peaks and dips in $\omega_{\rm I}$ are caused by mode crossings.
Large peaks in $\omega_{\rm I}$ can be interpreted as couplings with strongly damped modes 
associated with higher radial orders and higher $\lambda_{k'm}$
having $k'=2,~4,\cdots$.

Since $|\bar\omega_{\rm I}|$ of the $\ell=1$ ($\lambda_{0\pm1}$) $g$-modes is much smaller than
that of the $\ell = 3$ ($\lambda_{2\pm1}$) $g$-modes, zoomed diagrams of the former case
are given in Fig.~\ref{fig:omegaOmegak0}.
For the $\ell=1$ retrograde modes (panel (b)) $\bar\omega_{\rm I}$ 
behaves similarly to $\ell \ge 3$ cases with numerous high peaks.
This is reasonable because the frequency $\bar\omega_{\rm R}$ of 
a retrograde $\ell=1$ mode increase with $\bar\Omega$ 
similarly to the modes of $\ell\ge3$ so that 
similar mode couplings would occur. 
On the other hand,
the stability of {\it prograde} $\ell=1$ ($\lambda_{0-1}$) modes behaves differently;
for $\bar\Omega \ga 0.2$, sharp peaks are replaced with broad bumps which
become weak as $\bar\Omega$ increases, and those unstable modes stay unstable
even at large $\bar\Omega$.
This property is common for prograde sectoral modes 
and consistent with the results in Figs.~\ref{fig:grow4A} and
\ref{fig:grow4D} 
which show prograde sectoral modes being excited even at
a rapid rotation.

\begin{figure}
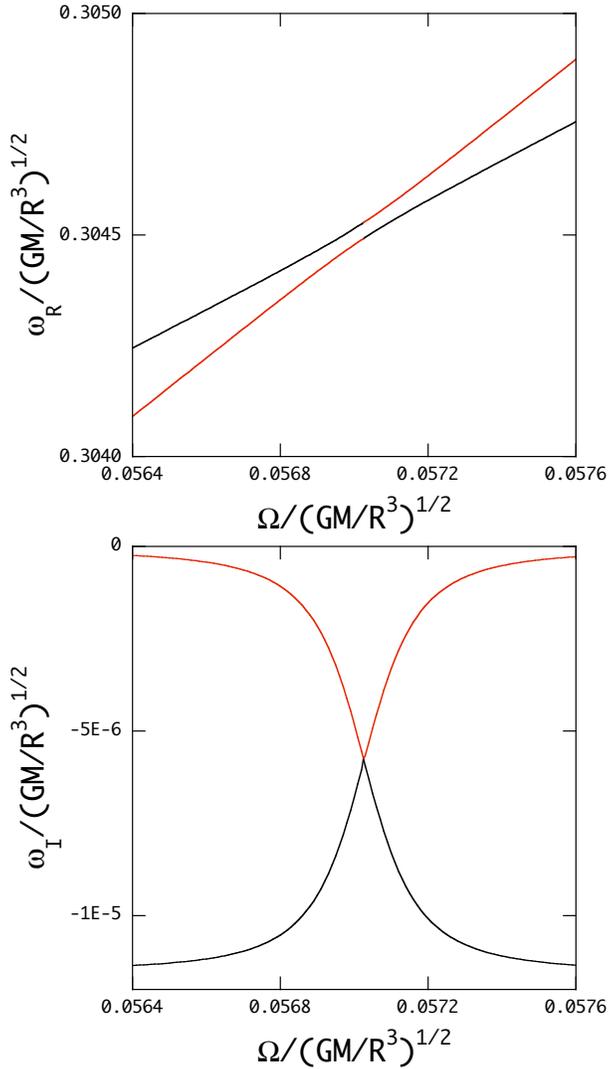

\resizebox{0.45\textwidth}{!}{
\includegraphics{f9a.epsi}}
\resizebox{0.45\textwidth}{!}{
\includegraphics{f9b.epsi}}
\caption{Mode crossing between $\ell=m=1$ $g_{9}$ mode (red line) 
and $\ell=3, m=1$ $g_{22}$ mode (black line).
The real part makes an avoided crossing, where the mode property is exchanged
at the closest approach.
}
\label{fig:couple922}
\end{figure}

Let us now look into two crossings which occur between modes of $(\ell,m) = (1,1)$ and
$(3,1)$ (or between $\lambda_{01}$ and $\lambda_{21}$) in detail.
Fig.~\ref{fig:couple922} shows a zoomed diagram around a crossing at 
$\bar{\Omega}\approx0.06$
which is recognized as connected small peak and dip in Fig.~\ref{fig:omegaOmega} (panel d).
The crossing occurs between the modes $g_{9}$ of $(\ell,m) = (1,1)$ (red line) 
and $g_{22}$ of $(\ell,m) = (3,1)$ (black line).
As this figure shows, the real parts of the eigenfrequencies make an avoided crossing, 
while the imaginary parts make a true crossing.
At the crossing the properties of the modes are interchanged.

\begin{figure}
\resizebox{0.45\textwidth}{!}{
\includegraphics{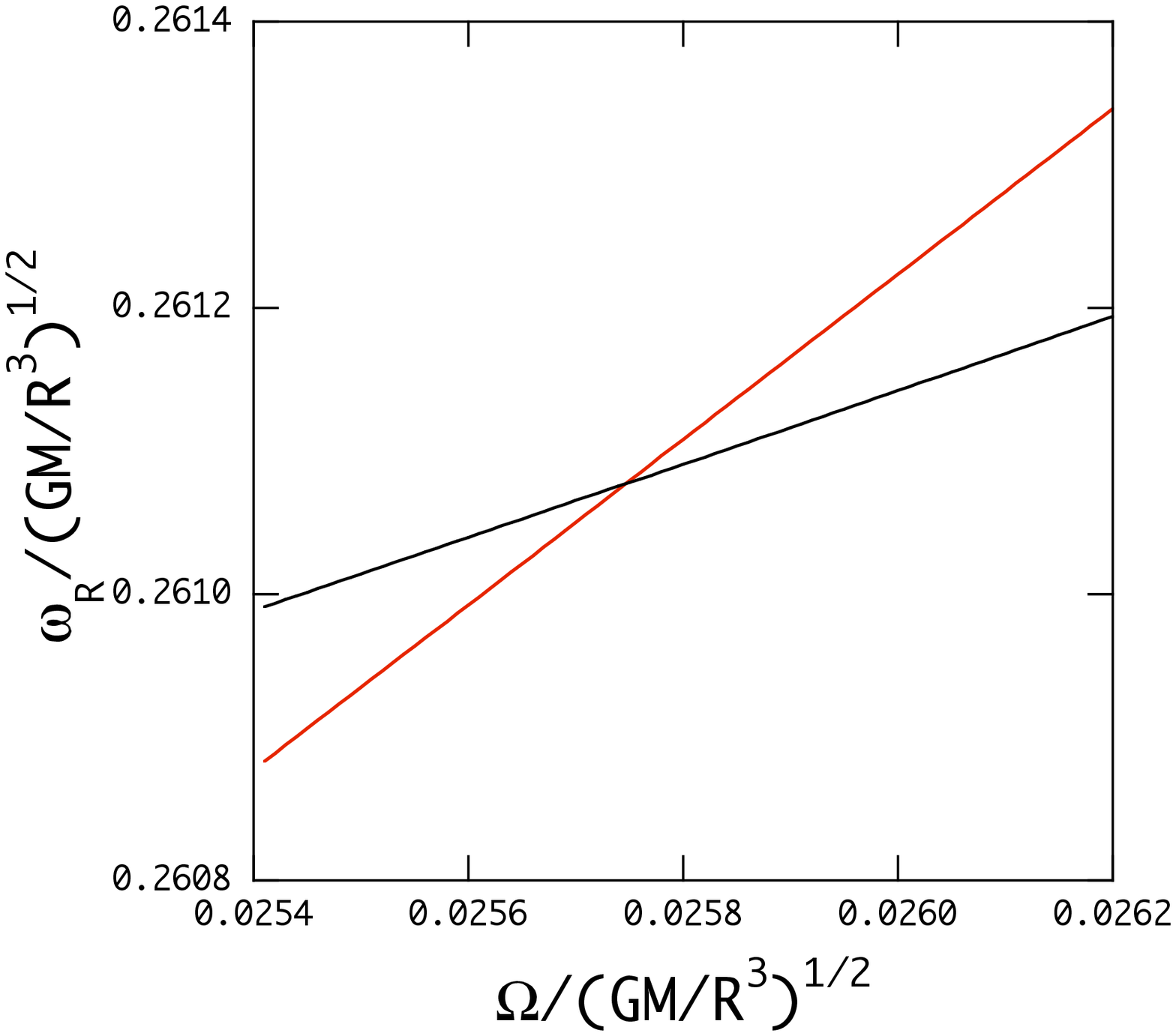}}
\resizebox{0.45\textwidth}{!}{
\includegraphics{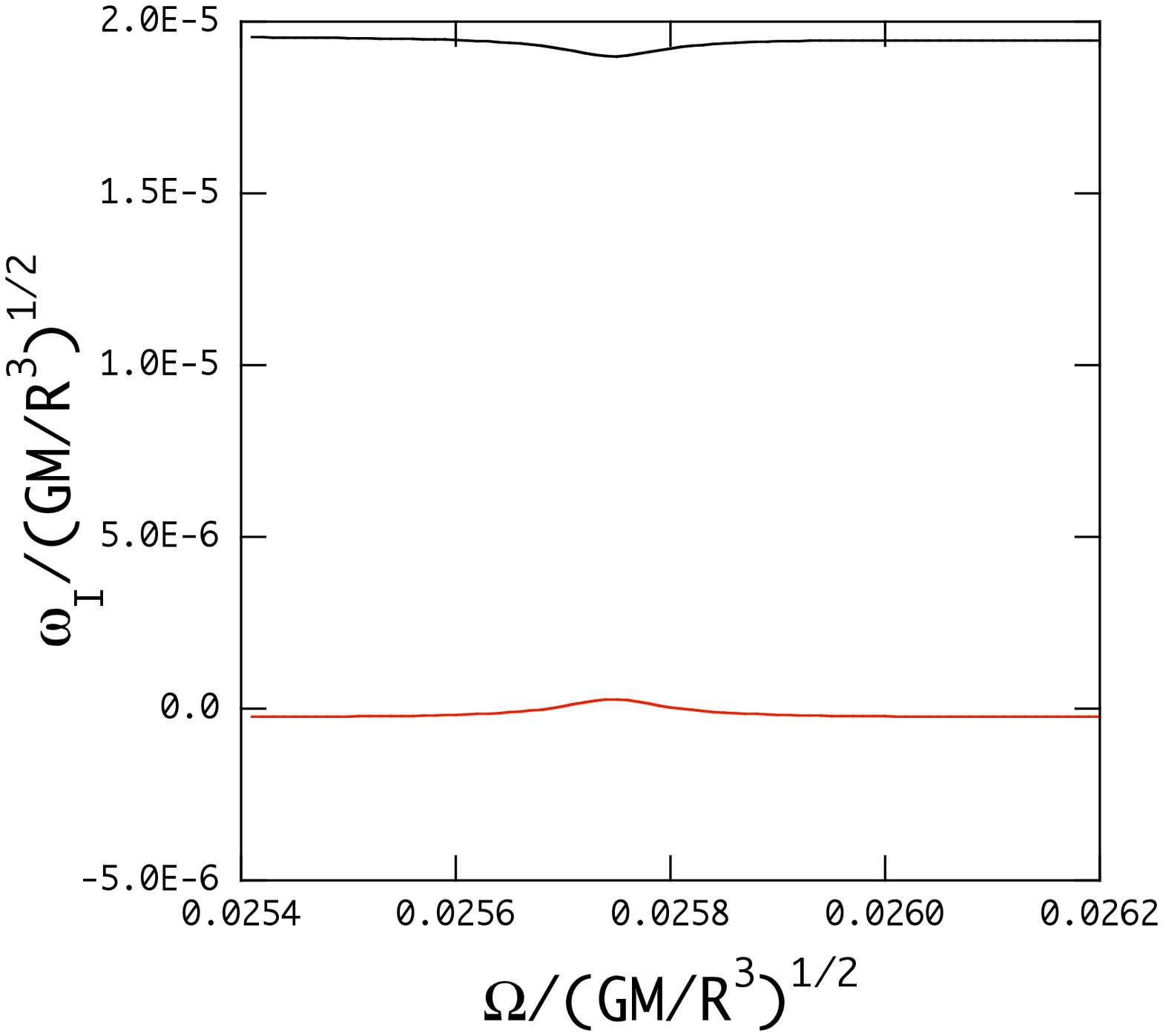}}
\caption{Mode crossing between $(\ell,m)=(1,1)$ $g_{10}$ (red line) and 
$(\ell,m)=(3,1)$ $g_{25}$ (black line) modes.}
\label{fig:couple1025}
\end{figure}

Fig.~\ref{fig:couple1025} shows another example of mode crossings which occurs 
between $g_{10}$-mode of $(\ell,m)=(1,1)$ (red line)
and $g_{25}$-mode of $(\ell,m)=(3,1)$ (black line) at $\bar\Omega\approx0.026$. 
The real parts of the frequencies appear to make a true crossing, while 
the imaginary parts show only a very small dip and bump.
This crossing may be understand as a crossing with an interaction much weaker
than the case of Fig.~\ref{fig:couple922}. 
The weak interaction reduces the distance at the closest encounter 
of $\omega_{\rm R}$s to zero and causes only very small effects on the stability of the modes.

\subsection{Asymptotic Analysis}
\subsubsection{Coupling coefficient}
It is helpful to discuss the mode crossing phenomena by using 
an asymptotic method based on the traditional approximation.
Under the traditional approximation, the oscillation modes of a rotating star 
are separated into independent modes associated 
with different $\lambda_{km}$s.
The traditional approximation works well for $g$-modes except when the 
frequencies of two modes with the same $m$ and the same (even or odd) 
parity become very close to each other. 
Under the traditional approximation these two modes are independent,
but in reality the term $-\Omega\sin\theta\pmb{e}_\theta$, neglected 
in the approximation, brings about mode coupling between them.

Here, we extend the asymptotic analysis for adiabatic oscillations 
developed by \citet{Lee89} 
to the case of non-adiabatic oscillations.
Note that no effects of the centrifugal force are included in the following 
asymptotic treatment, and only the effects of the Coriolis force are considered.
\citet{Lee89} have derived a dispersion relation which should be satisfied
by adiabatic $g$-modes associated with $\lambda_{jm}$
and $\lambda_{km}$;  
\be
\tan\Psi_j \tan\Psi_k=\epsilon_{jk},
\label{eq:disp}
\ee
where 
\be
\Psi_j=-\int_{x_c}^{1}k_j{dx}+{n_e\over 2}\pi,
\label{eq:psi}
\ee
$x_c$ denotes the outer boundary of the convective core,
$n_e$ is the effective polytropic index at the surface, and
\be
k_j=
{\sqrt{\lambda_{jm}}\over x}{\bar N\over\bar\omega}.
\label{eq:kj}
\ee
The right hand side of equation (\ref{eq:disp}), $\epsilon_{jk}$, represents
the effect of coupling between the two $g$-modes.
We call it ``coupling coefficient''. 
Note that if the coupling were absent (i.e., $\epsilon_{jk}=0$), 
equation (\ref{eq:disp}) would be 
reduced to $\tan\Psi_j=0$ and $\tan\Psi_k=0$, 
which give essentially the same expressions for $\bar{\omega}_j$
and $\bar{\omega}_k$ as equation (\ref{eq:freq}).

The coupling coefficient $\epsilon_{jk}$ arises from a deviation from  
the traditional approximation around a crossing between $g$-modes associated 
with $\lambda_{jm}$ and $\lambda_{km}$.
\citet{Lee89} derived the expression  
\be
\begin{array}{l}\displaystyle
\epsilon_{jk}=\left[\int_{x_c}^1{dx\over x}\left(
\left|{\lambda_{jm}\over\lambda_{km}}\right|^{1/4}G_{jk}\cos\chi^j\sin\chi^k \right.\right. 
\cr \hspace{0.3\columnwidth} \displaystyle \left.\left.
+ ~\left|{\lambda_{km}\over\lambda_{jm}}\right|^{1/4}G_{kj}\cos\chi^k\sin\chi^j \right)
\right]^2,
\end{array}
\label{eq:epsilon}
\ee
where 
\be
\chi^j=-\int_{x_c}^xk_jdx+{n_e\over 2}\pi
\ee
and $G_{jk}$  
is a matrix element defined in \citet{Lee89}, 
which consists of terms conflicting with the traditional approximation
and 
brings about coupling between the two modes.
The matrix element $G_{jk}$ is proportional to $\nu$ so that
$\epsilon_{jk}\rightarrow 0$ as $\nu\rightarrow 0$.
This corresponds to the fact that modes with different $\ell$s 
are independent in a non-rotating star.

\begin{figure}
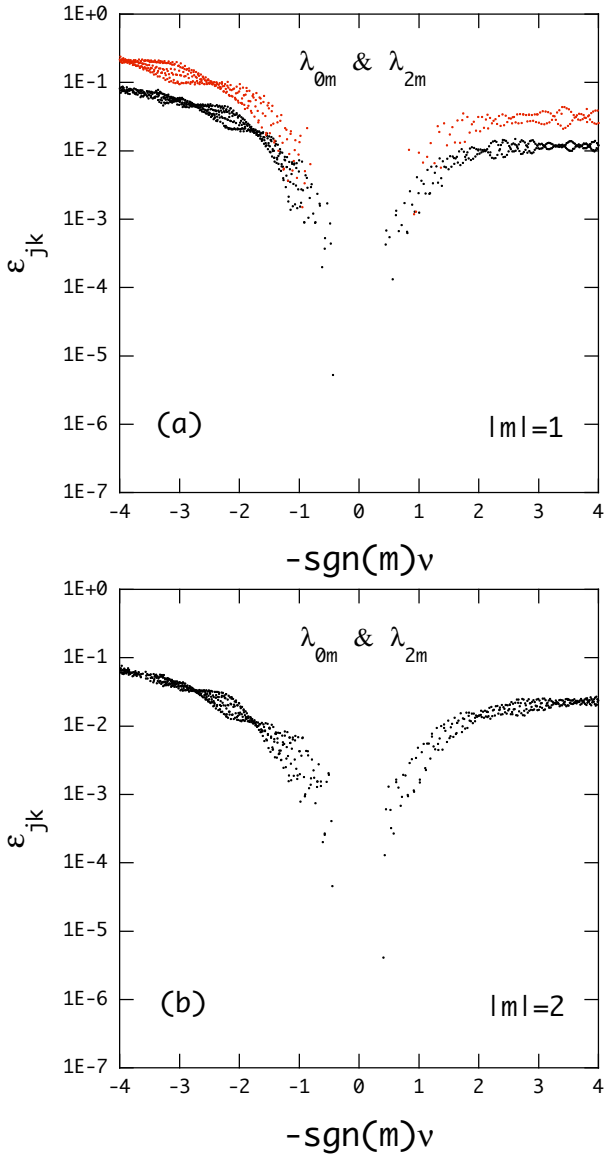

\resizebox{0.45\textwidth}{!}{
\includegraphics{f11a.epsi}}
\resizebox{0.45\textwidth}{!}{
\includegraphics{f11b.epsi}}
\caption{Coupling coefficient $\epsilon_{jk}$ between even $g$-modes 
associated with $\lambda_{0m}$
and $\lambda_{2m}$ for $|m|=1$ in panel (a) and for $|m|=2$ in panel (b)
{\bf as a function of $\nu=2\Omega/\omega$}.
Here, we have used the $4M_\odot$ ZAMS model, and assuming $\bar\Omega=0.2$ 
{\bf (black dots) and $\bar\Omega=0.4$ (red dots)} we have calculated $g$-modes
in the frequency range between $\bar\omega=0.1$ and 1.}
\label{fig:eps02}
\end{figure}

\begin{figure}
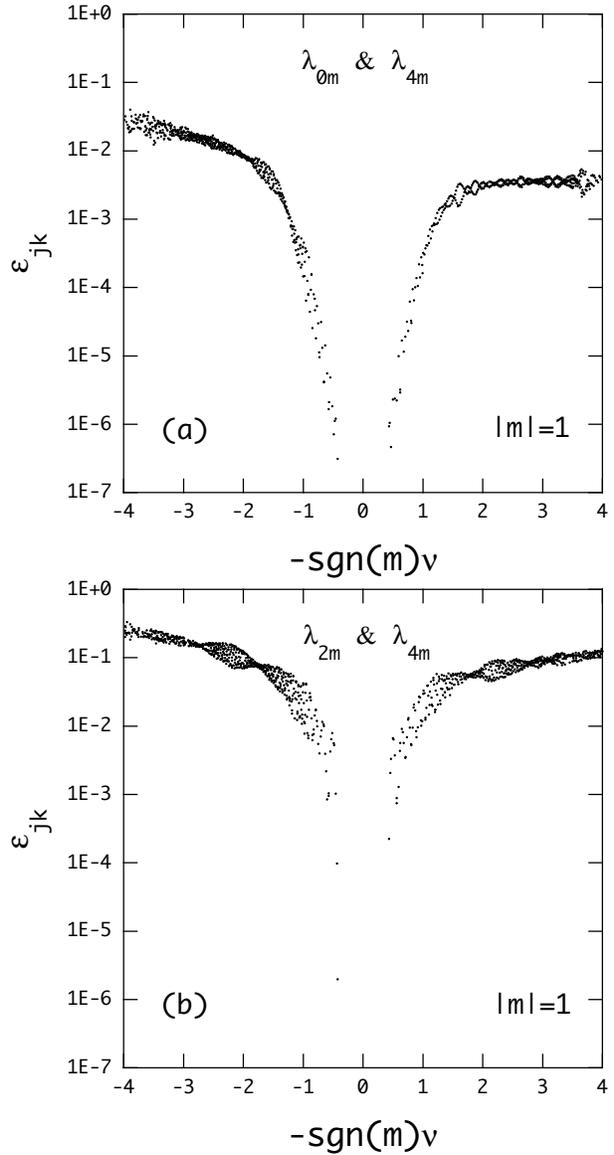

\resizebox{0.45\textwidth}{!}{
\includegraphics{f12a.epsi}}
\resizebox{0.45\textwidth}{!}{
\includegraphics{f12b.epsi}}
\caption{Coupling coefficient $\epsilon_{jk}$ between even $|m|=1$ $g$-modes associated with $\lambda_{0m}$
and $\lambda_{4m}$ in panel (a) and between those associated with $\lambda_{2m}$
and $\lambda_{4m}$ in panel (b).
Here, we have used the $4M_\odot$ ZAMS model, and for $\bar\Omega=0.2$ we have calculated $g$-modes in the frequency range between $\bar\omega=0.1$ and 1.}
\label{fig:eps04}
\end{figure}

%
From equation (\ref{eq:epsilon}), we can calculate $\epsilon_{jk}$
as a function of $\nu$ (or $\bar{\omega}$) for given $m$, $\Omega$, 
and $(j,k)$. However, only discrete values of $\nu$ satisfy the dispersion
relation in equation (\ref{eq:disp}). 
We obtained frequencies satisfying the dispersion relation by solving
equation (\ref{eq:disp}) numerically with Newton's method in the frequency
range between $\bar\omega=0.1$ and $\bar\omega=1$ 
for the $4M_\odot$ ZAMS model. 
The values of $\epsilon_{jk}$ at the discrete frequencies thus obtained 
are shown in Fig.~\ref{fig:eps02} for $(j,k)=(0,2)$, and 
in Fig.~\ref{fig:eps04} for $(j,k)=(0,4)$ and (2,4). 
Each point corresponds to a $g$-mode belonging to $\lambda_{j,m}$ or
$\lambda_{k,m}$.
Since $\lambda_{j,m}$ (and/or $\lambda_{k,m}$) increases rapidly with $\nu$
(except for the case of $\lambda_{0m}$ with $m<0$; Fig.~\ref{fig:lambda}),
a large number of $g$-mode frequencies enter into the range
$0.1\le\bar\omega \le 1$ (cf. eq.~\ref{eq:freq}).   
%
%

Fig.~\ref{fig:eps02} shows
the coupling coefficient $\epsilon_{jk}$ between 
even $g$-modes associated  with $\lambda_{0m}$ ($\ell=|m|$) 
and $\lambda_{2m}$ ($\ell=|m|+2$)
for $|m|=1$ (panel a) and $|m|=2$ (panel b) as functions of 
$\nu$. 
The coupling coefficients increase rapidly as the parameter $\nu$ increases,
which is reasonable because without rotation two modes 
with different $\ell$s are independent from each other.
For prograde ($m<0$) modes, the coefficients level off for $\nu\ga 2$ so that 
the coupling coefficients of prograde modes are appreciably smaller than 
those of retrograde modes when $\nu$ is sufficiently large.
 
Figure~\ref{fig:eps04}, shows coupling coefficients between 
$g$-modes associated with $\lambda_{0\pm1}$ ($\ell=1$) 
and $\lambda_{4\pm1}$ ($\ell=5$) 
in panel (a), and between those associated with $\lambda_{2m}$ ($\ell=3$) and 
$\lambda_{4m}$ ($\ell=5$) in panel (b).
Obviously, the latter is much larger than the former at a given $\nu$.

Fig.~\ref{fig:eps02}a shows that coupling coefficients are larger 
if $\bar\Omega=0.4$ (red dots) is used rather than 0.2 (black dots),
indicating coupling is stronger in a rapid rotator for a given value
of $\nu$.

We also found that the coupling coefficients for the $5M_\odot$ 
ZAMS model (not shown)
are similar to those of $4M_\odot$ ZAMS model shown in the figures.

\subsubsection{Quasi-adiabatic extension}
In order to study mode couplings between nonadiabatic $g$-modes with the dispersion relation,
it is necessary to extend it by including non-adiabatic effects. 
We modify the wavenumber $k_j$ 
in the quasi-adiabatic approximation as 
\be
k_{j}={\sqrt{\lambda_{jm}}\over x}{\bar{N}\over\bar\omega}
\sqrt{1+{\rm i}\beta_j{\delta_T\left(\nabla_{ad}-\nabla\right)
\over\delta_T\left(\nabla_{\rm ad}-\nabla\right)+\nabla_\mu}}
\label{eq:kjna}
\ee
\citep{Lee85}, 
where the oscillation frequency $\bar\omega$ is now regarded as a complex number, and
\be
\beta_j=\lambda_{jm}{\bar{N}^2\over\bar\omega^2}
{1\over\bar\omega c_2V\nabla}, \quad
c_2={4\pi r^3\rho Tc_p\over L_r}\sqrt{GM\over R^3},
\ee
$\delta_T=-(\partial\ln\rho/\partial\ln T)_p$,
$\nabla_{\rm ad}=(\partial\ln T/\partial\ln p)_{\rm ad}$, 
$\nabla=d\ln T/d\ln p$, $\nabla_\mu=d\ln\mu/d\ln p$, $V=-d\ln p/d\ln r$,
and $L_r$ is the radiative luminosity 
and $c_p$ is the specific heat at constant pressure.
Here, for simplicity we have considered only radiative damping as 
the non-adiabatic effect.
We note that if we define the local thermal time scale $\tau_{th}$ as
$\tau_{th}={4\pi r^2H_p\rho Tc_p/ L_r}$ with $H_p=-dr/d\ln p$ 
being the pressure scale-height,
we have $\bar\omega c_2/V=2\pi{\tau_{th}/ P}$
with $P=2\pi/\omega$ being the oscillation period, 
and the term containing $\beta_j$, 
which represents the radiative damping effect, is proportional to $P/\tau_{th}$.
For the quasi-adiabatic approximation to be valid, 
we take account of the non-adiabatic contribution
only in the region where $P/\tau_{th}\ltsim 0.01$.
This criterion, however, is somewhat ambiguous since we ignore destabilizing 
contributions near the surface where $P/\tau_{th}\sim 1$.

Replacing the radial wavenumber in equation~(\ref{eq:kj}) 
by the complex wavenumber given in equation~(\ref{eq:kjna}), 
we solve the dispersion relation in equation~(\ref{eq:disp}) 
to obtain complex frequencies $\bar\omega$ as functions of $\bar\Omega$,
where we use $n_e=3.525$.
Note that since we include only radiative damping as the non-adiabatic 
contribution, 
we obtain only stable modes having positive $\omega_{\rm I}$.

\begin{figure}
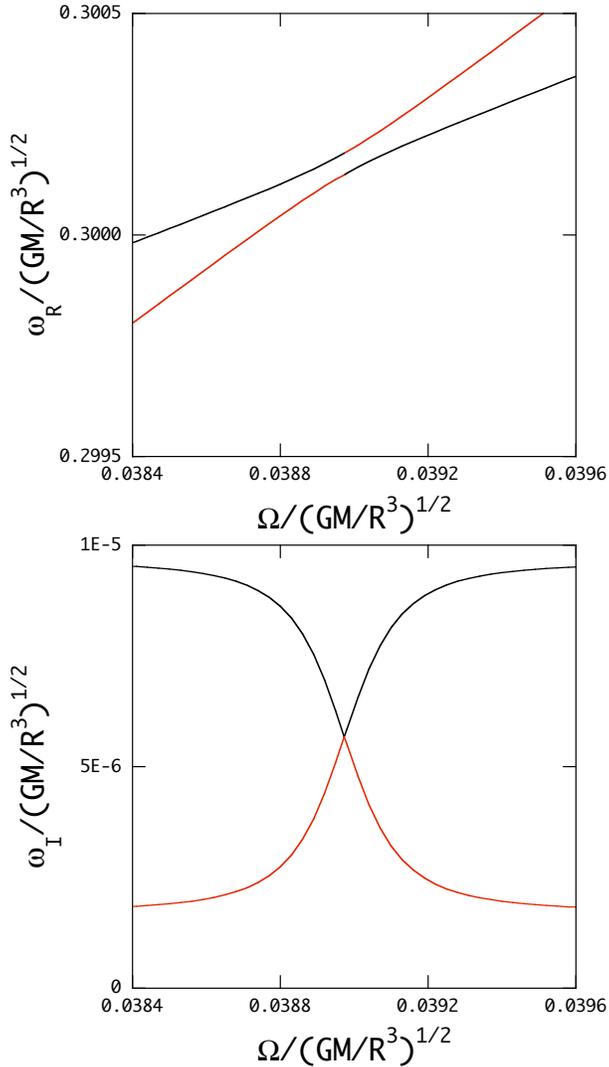

\resizebox{0.45\textwidth}{!}{
\includegraphics{f13a.epsi}}
\resizebox{0.45\textwidth}{!}{
\includegraphics{f13b.epsi}}
\caption{Complex $\bar\omega$ of even retrograde (m=1) $g$-modes with $\ell=1$ and $\ell=3$
calculated as solutions to the dispersion relation, eq.~(\ref{eq:disp}),
for the $4M_\odot$ ZAMS model.
To be compared with Fig.~\ref{fig:couple922}.}
\label{fig:dispavcr}
\end{figure}

\begin{figure}
\resizebox{0.45\textwidth}{!}{
\includegraphics{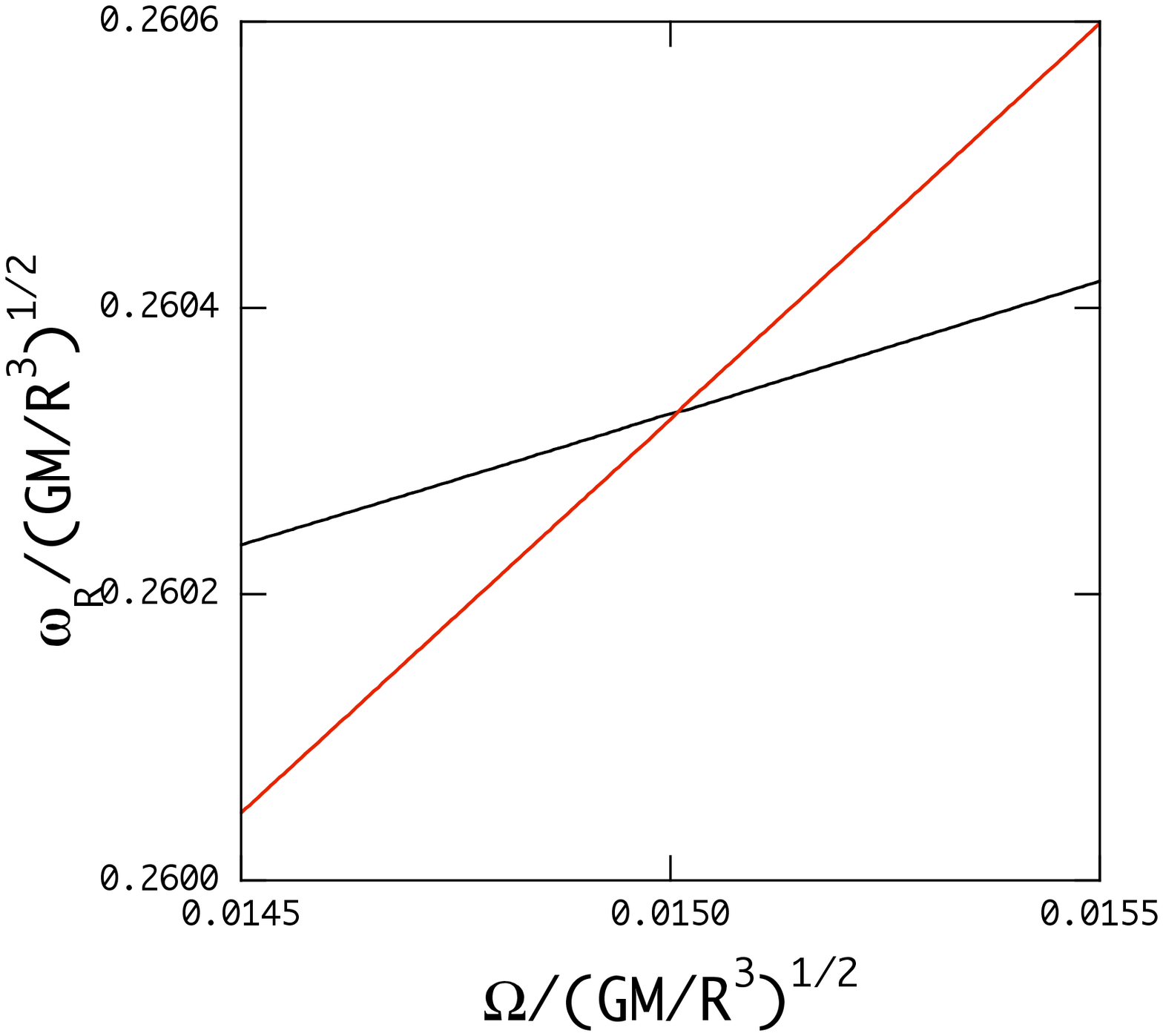}}
\resizebox{0.45\textwidth}{!}{
\includegraphics{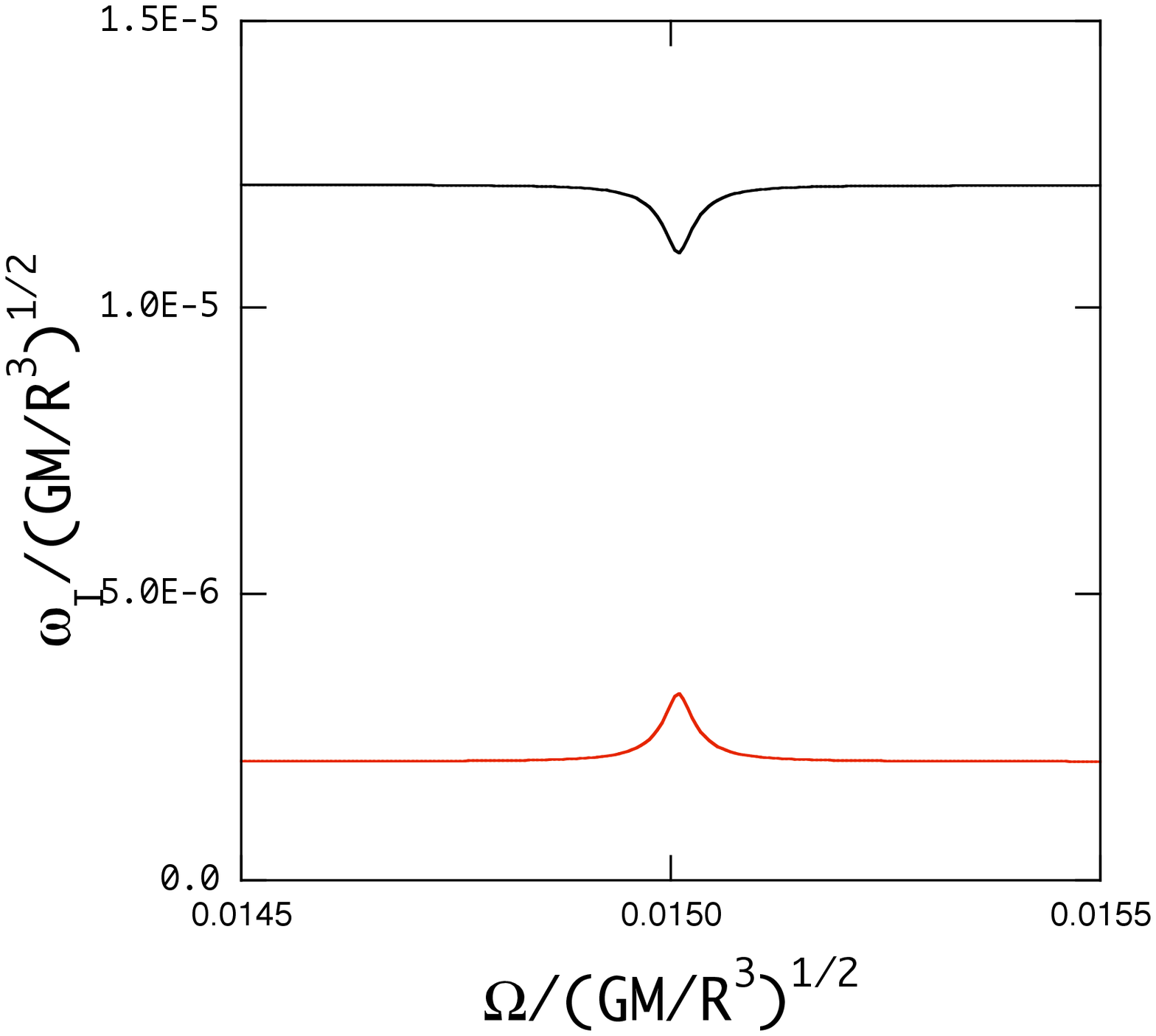}}
\caption{Complex $\bar\omega$ of even retrograde (m=1) $g$-modes with $\ell=1$ and $\ell=3$
calculated as solutions to the dispersion relation, eq.~(\ref{eq:disp}), 
for the $4M_\odot$ ZAMS model.
To be compared with Fig.~\ref{fig:couple1025}.}
\label{fig:dispcr}
\end{figure}

\begin{figure}
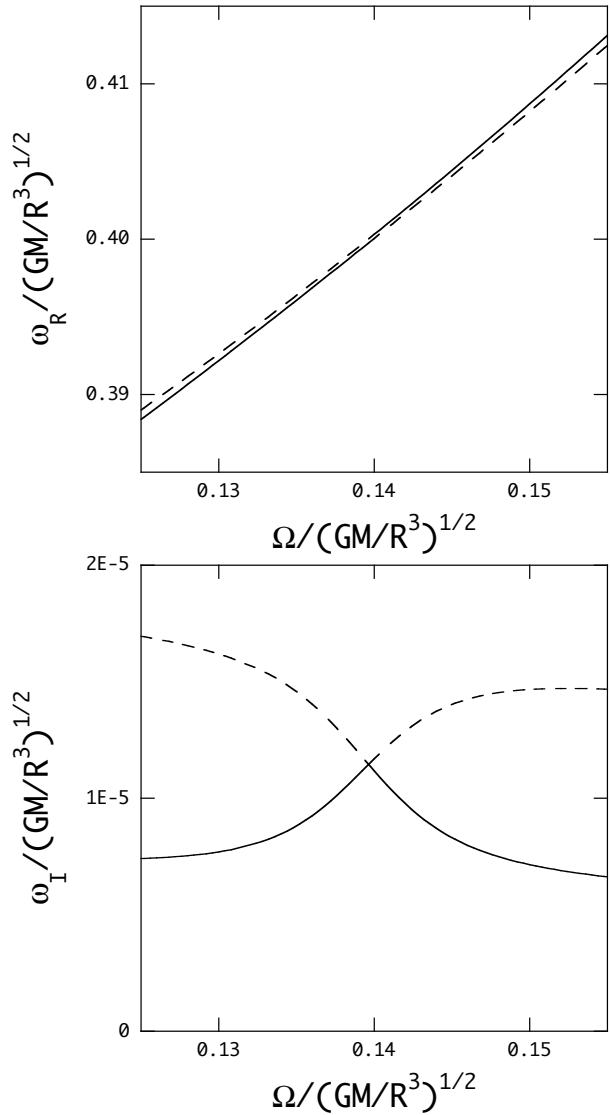

\resizebox{0.45\textwidth}{!}{
\includegraphics{f15a.epsi}}
\resizebox{0.45\textwidth}{!}{
\includegraphics{f15b.epsi}}
\caption{Complex $\bar\omega$ of even retrograde (m=1) $g$-modes with $\ell=3$ (solid line) and $\ell=5$
(dashed line) calculated as solutions to the dispersion relation for the $4M_\odot$ ZAMS model.}
\label{fig:disp35}
\end{figure}

Figs.~\ref{fig:dispavcr}--\ref{fig:disp35} show three examples of mode crossings 
calculated using the dispersion relation in equation~(\ref{eq:disp}). 
Fig.~\ref{fig:dispavcr} corresponds to the mode crossing in Fig.~\ref{fig:couple922},
in which the real parts of the eigenfrequencies make an avoided crossing
and the imaginary parts make a true crossing causing a large peak and dip. 
Fig.~\ref{fig:dispcr} corresponds to that in Fig.~\ref{fig:couple1025},
in which the real parts of the eigenfrequencies appear to make a true crossing and the
imaginary parts show a small dip and bump. 
The coupling coefficient $\epsilon_{jk}$ for the case of Fig.~\ref{fig:dispavcr} 
is of the order of $10^{-4}$, while for the case of Fig.~\ref{fig:dispcr} it is
of the order of $10^{-6}$, supporting the previous conjecture that 
a weak interaction leads to such a crossing as seen in Fig.~\ref{fig:couple1025}.
The value of $\bar\Omega$ at which
the mode crossing takes place does not necessarily agree well between 
the asymptotic method and the pulsation calculation. 
A possible reason
for the disagreement may be attributable to neglecting the effect of 
the centrifugal force in the asymptotic treatment.
%
%
The derivatives $d\omega_R/d\Omega$ for low frequency $g$-modes are
different between the cases with and without the effects of 
the centrifugal force on the modes, and
a slight difference in the derivatives may cause a large difference 
in the locations of $g$-mode crossing between 
the two cases having similar gradients  $d\omega_R/d\Omega$.
%
%

Fig.~\ref{fig:disp35} shows 
a mode crossing between two
$g$-modes associated with $\lambda_{2m}$ and $\lambda_{4m}$ for $m=1$ which
has a broad interaction range in $\bar\Omega$. 
The broad interaction is possible if the oscillation frequencies of the two modes
have similar dependence on the rotation frequency (i.e., similar $d\omega_{\rm R}/d\Omega$)
around the crossing,
and the coupling coefficient is sufficiently large.
The latter requirement is obviously met for the crossings between
$\lambda_{2m}$ and $\lambda_{4m}$ $g$-modes (see panel (b) of Fig.~\ref{fig:eps04}).

The frequency dependence on the rotation frequency can be estimated from
equation~(\ref{eq:freq}).
Since  $\lambda_{jm}$ is proportional to $\nu^2$ for sufficiently
large $\nu$ (except for prograde sectoral modes with $\lambda_{0m}$ and $m<0$),
then, $\omega_{\rm R}$ becomes proportional to $\sqrt{2\Omega/n}$ independent of $j$
(or $\ell$). Therefore, most of the $\bar\omega_{\rm R}$-$\bar{\Omega}$ curves
become closer to parallel to each other so that broad interactions such as the one shown in
Fig.~\ref{fig:disp35} would occur frequently as $\bar{\Omega}$ increases.
This explains broad and large peaks of $\bar\omega_{\rm I}$ for tesseral modes
with $\ell > |m|$ in the panels (b) and (d) of Fig.~\ref{fig:omegaOmega} and
for {\it retrograde} sectoral modes with $\ell = m$ with $m>0$ in the panel (b) of
Fig.~\ref{fig:omegaOmegak0}. 

The {\it prograde sectoral} modes are exceptional. Since $\lambda_{0m}$ with
$m<0$ is nearly constant for $\nu \ga 1$, they experience only narrow interactions
with tesseral modes.
This explains why peaks for the prograde sectoral modes seen in the panel (a) of
Fig.~\ref{fig:omegaOmegak0} are narrower than those in the other cases. 
The bumps seen for $\bar\Omega \ga 0.2$ 
can be understood as swarms of numerous weak and narrow interactions.

\section{conclusion}

We studied the stability of low degree $g$-modes 
in uniformly rotating main-sequence stars with masses of $4M_\odot$ and $5M_\odot$,
by taking into account the effects of both the 
Coriolis force and the centrifugal force, 
using the method of calculation given by \citet{Lee95}. 
From the analysis treating $\bar{\Omega}$ as a small parameter we found
that a slow rotation has destabilizing (stabilizing) effect on high 
radial-order retrograde (prograde) $g$-modes,
although the effects for relatively low-order modes are very small or absent.
This effect can be understood from the relation between period change due
to rotation and the optimal period for the kappa-mechanism at the Fe-opacity bump.

Calculating eigenfrequencies of low degree $g$-modes at selected rotation frequencies,
we found that,
on the other hand, rapid rotation tends to stabilize high radial-order retrograde $g$-modes, 
and the stabilizing effect appears stronger for less massive stars with lower 
effective temperatures. 

Obtaining the eigenfrequency of a $g$-mode of a degree $\ell$ 
(associated with a $\lambda_{km}$) 
as a function of the rotation frequency $\Omega$, we found that it 
experiences mode crossings with $g$-modes of different $\ell'$s (associated 
with different $\lambda_{k'm}$s).
At a mode crossing between two modes with the same $m$ and parity
(even or odd) a coupling occurs because these two modes are not independent
in a rotating star.
If an unstable mode crosses a damped mode, the former can be stabilized around
the crossing, or vice versa.
At a large rotation frequency, low $\ell$ $g$-modes, retrograde ones in particular,
tend to be stabilized by mode couplings with larger $\ell$ modes which are strongly damped.
The {\it prograde sectoral} modes ($\ell=-m$ and $m<0$) in rapidly rotating stars
receive exceptionally weak damping effects from mode couplings, because the 
$\Omega$ dependence of their frequencies is very different from
the other modes.

We derived a dispersion relation useful to study mode coupling 
including the radiative damping effect under the quasi-adiabatic approximation.
In the dispersion relation the strength of a mode coupling is controlled by 
a coupling coefficient, $\epsilon_{jk}$ between two $g$-modes belonging to
$\lambda_{jm}$ and $\lambda_{km}$ ($j>k$); a larger $\epsilon_{jk}$ leads to a stronger
coupling. 
The value of $\epsilon_{jk}$ is larger when $k\not=0$ and when  $1/\omega_{\rm R}$ and
$\Omega$ are larger, and it tends to be larger for retrograde modes than for prograde modes.   
By using the dispersion relation for various combinations of modes, we
could reproduce all types of mode couplings.

To include the effects of the rotational deformation of the equilibrium structure,
we employed the Chandrasekhar-Milne (C-M) expansion, where the deformation
is assumed to be proportional to $\Omega^2P_2(\cos\theta)$. 
This approximation would not be very accurate for a star rotating at a nearly critical
rate.
One way to avoid the problem would be to employ two-dimensional models, as
\citet{bal10} 
obtained adiabatic $g$-mode spectrum of two-dimensional rotating polytropes.
We hope that it will become possible in the near future to calculate the frequency spectrum and 
stability for $g$-modes of two-dimensional evolutionary models.

\end{document}